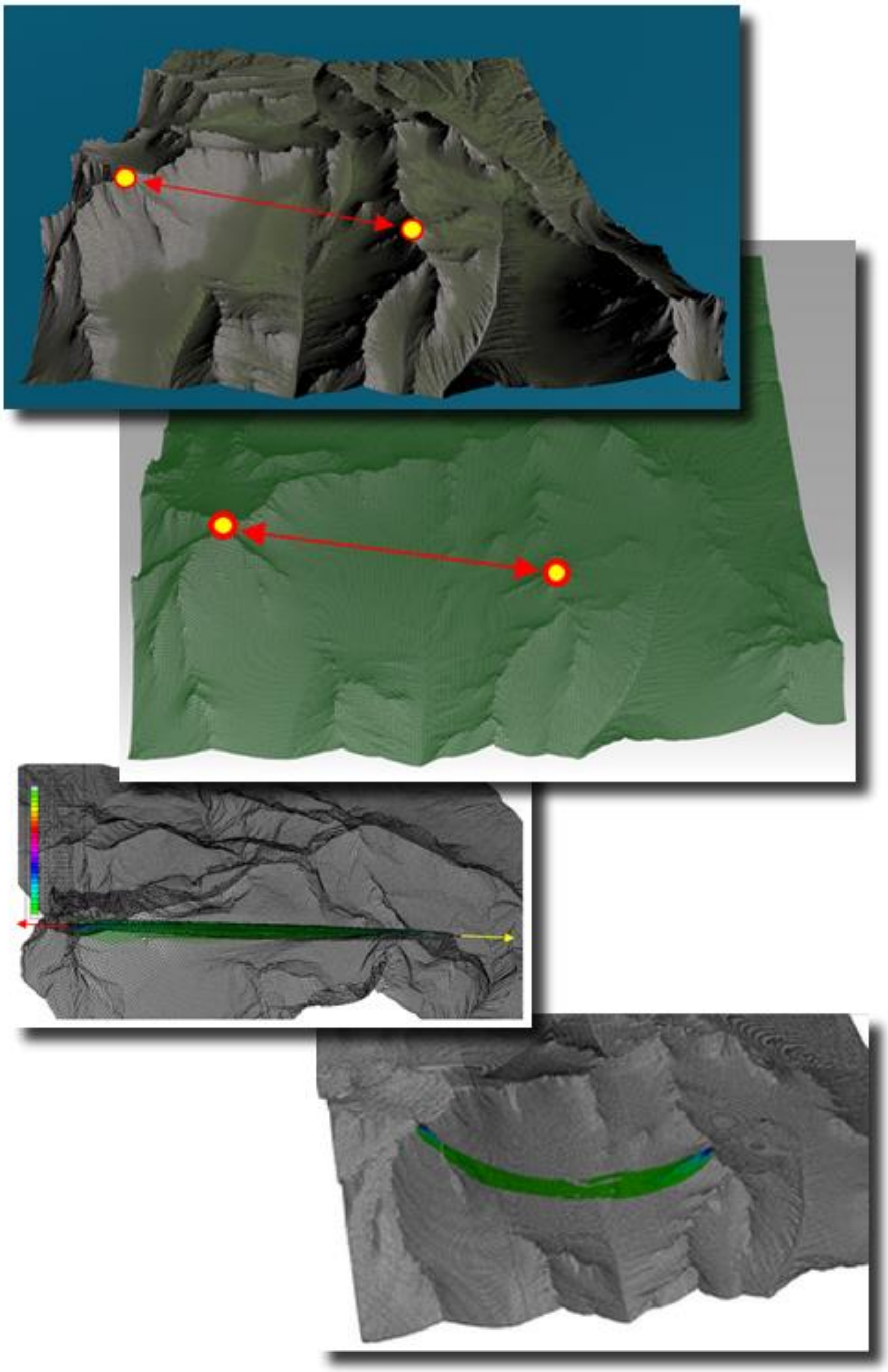

Gökhan Altıntaş

# Finding Shortest Path on a Terrain Surface by Using Finite Element Method

# Finding Shortest Path on a Terrain Surface by Using Finite Element Method

**Gökhan Altıntaş**

MANİSA CELAL BAYAR ÜNİVERSİTESİ İNŞAAT MÜHENDİSLİĞİ

Report Title

Finding Shortest Path on a Terrain Surface by Using Finite Element Method

| Author(s) | Report Identification Code |
|---|---|
| Gökhan Altıntaş | MCBUCIVILENG.R-2022.1 |

**Abstract**

The solution of the shortest path problem on a surface is not only a theoretical problem to be solved in the field of mathematics, but also problems that need to be solved in very different fields such as medicine, defense and construction technologies. When it comes to the land specific, solution algorithms for these problems are also of great importance in terms of determination of the shortest path in an open area where the road will pass in the field of civil engineering, or route determination of manned or unmanned vehicles for various logistic needs, especially in raw terrains.

In addition, path finding problems in the raw terrains are also important for manned and unmanned ground vehicles (UGV) used in the defense industry. Within the scope of this study, a method that can be used for instant route determinations within sight range or for route determinations covering wider areas is proposed. Although the examples presented within the scope of the study are land-based, the method can be applied to almost all problem types of similar nature. The approach used in the study can be briefly described as the mechanical analysis of a surface transformed into a structural load bearing system based on mechanical analogies.

In this approach, the determination of the shortest path connecting two points can be realized by following the stress-strain values that will occur by moving the points away from each other or by following a linear line that will be formed between two points during the mechanical analysis. If the proposed approach is to be carried out with multiple rigid body dynamics approaches instead of flexible bodies mechanics, it can be carried out easily and very quickly by determining the shortest path between two points or by tracking the forces. However, the proposed approach in this study is presented by simulating examples of flexible bodies using the finite element method. The approach used in this study is based on the approaches presented in Altintas [144].

Contact Addresses & Affiliations

Prof. Dr. Gökhan Altıntaş

Email(s): gokhanaltintas@gmail.com gokhan.altintas@cbu.edu.tr

Address: Manisa Celal Bayar University, Civil engineering Department., Engineering Faculty

B Blok, Şehit Prof. Dr. İlhan Varank Kampüsü 45140, Yunusemre Manisa Turkey

## Abstract

The solution of the shortest path problem on a surface is not only a theoretical problem to be solved in the field of mathematics, but also problems that need to be solved in very different fields such as medicine, defense and construction technologies. When it comes to the land specific, solution algorithms for these problems are also of great importance in terms of determination of the shortest path in an open area where the road will pass in the field of civil engineering, or route determination of manned or unmanned vehicles for various logistic needs, especially in raw terrains.

In addition, path finding problems in the raw terrains are also important for manned and unmanned ground vehicles (UGV) used in the defense industry. Within the scope of this study, a method that can be used for instant route determinations within sight range or for route determinations covering wider areas is proposed. Although the examples presented within the scope of the study are land-based, the method can be applied to almost all problem types of similar nature. The approach used in the study can be briefly described as the mechanical analysis of a surface transformed into a structural load bearing system based on mechanical analogies.

In this approach, the determination of the shortest path connecting two points can be realized by following the stress-strain values that will occur by moving the points away from each other or by following a linear line that will be formed between two points during the mechanical analysis. If the proposed approach is to be carried out with multiple rigid body dynamics approaches instead of flexible bodies mechanics, it can be carried out easily and very quickly by determining the shortest path between two points or by tracking the forces. However, the proposed approach in this study is presented by simulating examples of flexible bodies using the finite element method. The approach used in this study is based on the approaches presented in Altintas [144].

# TABLE OF CONTENTS

# LIST OF FIGURES

## Introduction

Determination of the shortest path on a surface is not only a mathematical problem, but also one of the important problems that need to be solved in areas different from each other. It is often used in determining the shortest path between two points in the field [1-41], which is one of the typical application areas for the solution of the shortest path problem on a surface. In addition, solving many problems with appropriate geometry in different fields such as electrical engineering [40, 41, 42-47] and medicine [41-52] are also among the problem types in this scope. However, in the literature, it is possible to say that the majority of the publications in the context of the application areas are the publications on the applications [29-32, 40, 41, 45, 52-57] and algorithm developments [34-39, 43-46, 58-81] of navigation problems.

It is possible to divide the surfaces of interest into two, mainly discretized and analytical surfaces, within the scope of the subject, where current studies continue without slowing down. Although the studies in the scope of analytical surfaces, which is one of the pioneering studies of the subject in mathematical context [82-93], still maintain their importance and actuality, studies dealing with discrete surfaces are increasingly taking place in the literature due to the support of developments in the fields of computational mathematics and engineering and their wide use in practice [94, 95]. In this context, it is possible to come across many studies in the current literature on the solutions of polyhedral surfaces [12-14, 96-107] and some of the studies within this scope have also focused on the exact solutions [108-112]. One of the biggest reasons why current studies are carried out with polyhedral surface definitions is that it is much more applicable to represent terrain and brain surface-like surface geometries, which are needed intensely for shortest path problem solutions, with polyhedral definitions, despite the difficulty of converting them as a whole to an analytically represented surface.

The issue is particularly important for navigation situations on unstructured, raw terrains. First of all, even if the surface itself is handled as a discrete surface instead of analytical definitions, a much more difficult problem has to be solved compared to the problem types on which road-dependent navigations are made.

The importance of solving such problems, in which there are much more possible routes between two points in terms of mathematics, emerges as an increasing need. The reason for this situation is closely related to the navigation needs of unmanned autonomous vehicles and robots as well as manned vehicles that move dependent on grounds [2-11, 15-25, 27, 28, 47, 60-81, 113-115, 116-118]. All logistics needs for commercial, agricultural or military purposes [116-118] can be considered in this context. Innovative methods, some of which are quite different from each other, constitute an important part of the current literature in the field of path planning [2-11, 47, 60-81, 115, 118-140]. The results obtained from these studies and the algorithms created are not only important for medical problems where CT and MRI data are used, and for path planning problems created by considering large areas where data such as GIS etc. are used, but also important for path planning problems where environmental conditions can change dynamically within the sight range. In addition, these studies are also necessary for the navigation needs of piloted or autonomous Rovers that are sent not only on Earth but also locations out of Earth [20-28].

In this study, approaches that can be easily applied with the already widely used FEA programs or the built-in physics engines of game engines are proposed. The approaches used can be compared to combinatorial approaches from various angles or surface flattening approaches [141-143]. However, mechanical analogies are based on different foundations from these approaches and thus offer a range of solutions in which new and different approaches are included. In the case of using the mechanics of flexible bodies, it is possible to reach the results quickly by using values such as strain, deformation, position and form, or by using criteria such as position, form or the values of forces with the approaches of connected rigid body dynamics or connected particle dynamics. It should be noted that mechanical analogies are not limited to those presented in this study, but can easily be further diversified.

The approaches presented in this study can be implemented with many mechanical analysis and simulation programs currently used in engineering and simulation fields. In this study, the results were obtained by using the analyzes made with the finite element method in the examples presented.

## Problem definition and solution approaching

The approach proposed in this study for the determination of the shortest path between two points on a surface is an adapted version of the approach proposed by Altintas [144,145] for the solution of optimization problems to be applied to surfaces. Before presenting visually supported applications about the presented approach, briefing about the preferred approach in the study will make the reasons for making some choices more understandable. Since mechanical analogies are essentially independent of mathematical techniques, the presented approach can be applied to surfaces that are defined as analytical or discrete. However, in the case of real-world problems, it is often not possible to express surfaces analytically. Expressing geometries with discrete elements and polyhedral surfaces, as in frequently encountered path finding problems in terrain, is a useful and adaptable approach to different problems. Although mechanical approaches are independent of numerical methods, special care should be taken not to include conditions arising from numerical methods as boundary conditions in the problem in some cases.

In order to apply the approach presented in this study appropriately to the problems, discrete surfaces were studied and the finite element method was used as a numerical method in the realization of mechanical simulations. Finite element method is a very suitable method for use in the mechanical modeling of discrete surfaces, as in this study. However, by transforming the surfaces into models of rigid body dynamics and connected particles dynamics with appropriate transformations, the models obtained can be simulated even by game engines. Since the subject can also be discussed from a topological point of view, in order to use a correct terminology, although the definition of "geodesic distance" is the same as the shortest path in most geometries, the term shortest path was used in the study due to the existence of exceptional cases [108].

The types of problems to be dealt with in this study are the determination of the shortest distance between two points on the same side of a surface geometry. The examples presented in the study are mainly about finding the distance between two points, which isolates a certain part of the land from the terrain. The two points in question can be two distant points on the terrain or two very close points within close region of sight, in both cases the suggested approach can be used. The proposed mechanical analogy is based on the principle of transforming the surface into a structure that can act like a fabric and obtaining the shortest path between them by moving the two points apart (Figure 1), depending on various mechanical criteria.

Although the above recipe is simple, there are many points to be considered in practice. Before addressing these issues, it may be useful to make a brief comment on Figure 1, which includes a simple example whose result can be easily predicted even without analysis, in order to understand the introductory example. In order to determine the shortest path between points A and B in Figure 1, a finite element model created in such a way that the surface can behave like a fabric is used to simulate the situation where the points can be moved apart from each other appropriately.

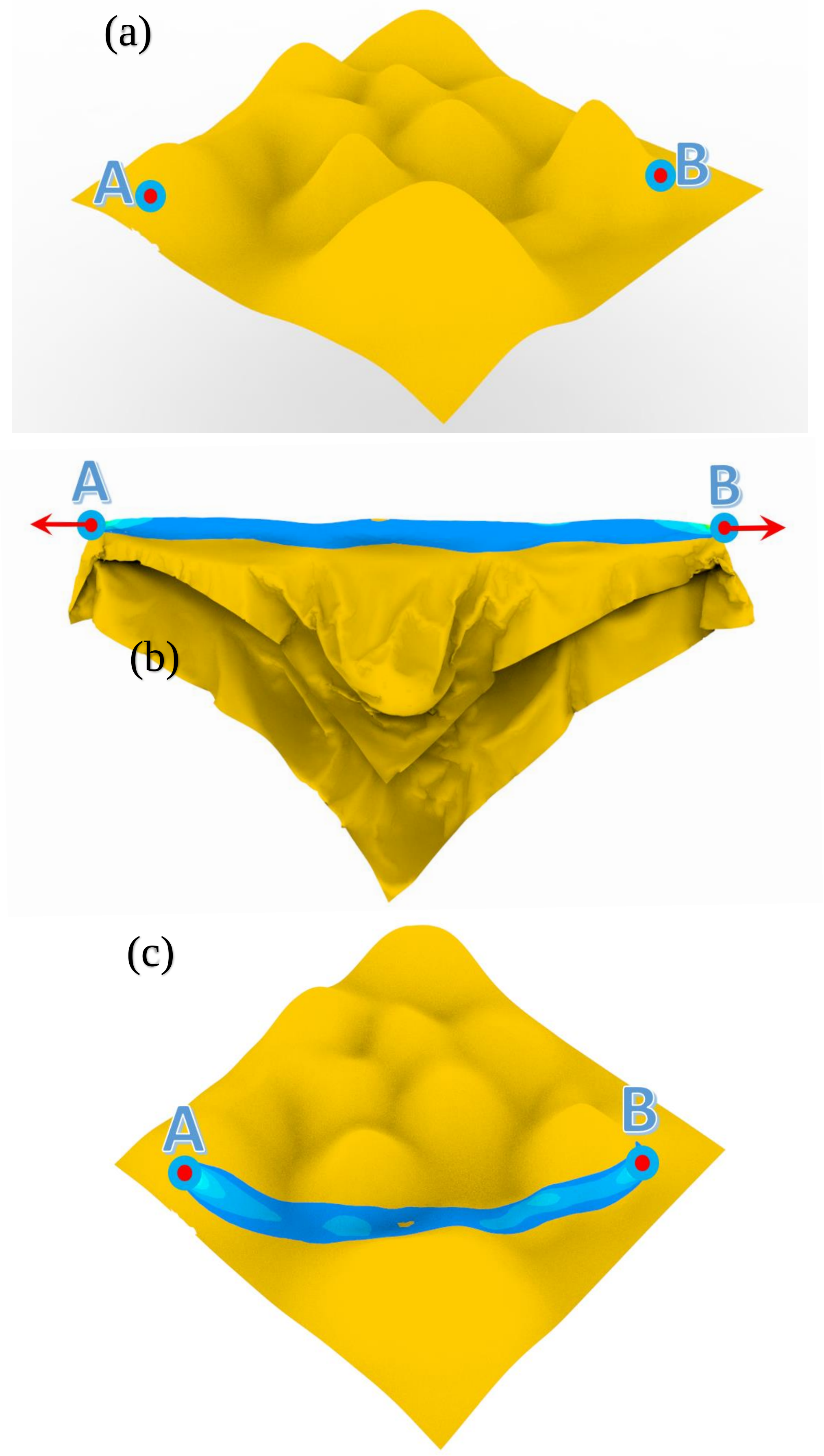

**Figure 1: (a)** Original terrain surface with nodes, **(b)** Stretched terrain surface and shortest path region (based on stresses), **(c)** Shortest path region on undeformed terrain surface

The result was obtained by following the changes in the stress, deformation and structural form that occurred during the simulation. In fact, it is possible to reach a result if only one of these evaluation criteria is used appropriately. The example, which was prepared for the visual intelligibility of the subject in a natural way, was used in the simulation, with properties such as gravity and interaction, which would normally be more appropriate not to be used. In the example in Figure 1, where gravity and interaction properties are used in terms of visual intelligibility, they do not have considerable effects on the solution set, but using these two properties in mechanical analogies may change the solution set and lead to erroneous results [146].

Due to the use of the finite element method as a numerical method, the results that can be obtained such as stress and deformation can be affected by properties such as materials and geometry. Although these outputs such as stress, deformation and displacement are evaluation criteria on the way to a solution, keeping them within certain limits prevents the emergence of erroneous solutions. The finite element method, which is the numerical method used in this study, was used in order to follow and comprehend the results of the mechanical analogy in the easiest way. Whether it is the best and feasible method of investigation may be a subject of separate criticism. However, it can be said that it is both possible and often faster and more effective to obtain solutions with rigid body dynamics-based approaches without the need for the use and management of outputs such as stress and deformation. The operations to be done in order to obtain the solution set as precisely as possible are not only related to getting rid of unnecessary properties such as interaction and gravity in analogy, but also closely related to other optional features of the approach to be used in the numerical method.

The element types used in this study and the selections related to the meshes are also examined in sample solutions because they have an effect on the results. It should be considered that an important part of the information to be given about the configurations of finite element meshes in this section are the arrangements that can be used in simulations based on rigid body dynamics.

## Element selection and meshes

Speaking of using FEM as a numerical method, it can be thought that the most accurate element type that can be used to represent a surface is the surface element types (shell, membrane, etc.). However, in the problem of interest, the geometric form that should be considered in the first plan for a path expression between two points is a line rather than a region to be presented by surface type elements. While using surface elements in finite element analysis, it is also possible to obtain a solution set that will form a set of linear lines, albeit indirectly, with post-processing operations. However, starting with the idea of finding a region from the beginning as a solution set carries the risk of getting away from certainty, but this method has its own advantages. For this reason, analyzes were made with different element types in the study in order for the practitioners to choose the most suitable approach for them.

The source of the surface presented in Figure 2(a) can be a point cloud and data from another source and in any case can be subjected to the triangularization process used in the calculations as seen in Figure 2(b). At this stage, the result of the problem of interest can be obtained with surface elements using the stretching process in mechanical analogy. In solutions obtained by using surface elements, it is possible to obtain regional solution sets as seen in Figure 1 based on parameters such as stress or deformation. With this approach, it can be possible to reduce the regions that make up the solution sets based on the numerical values of the parameters to be used in the evaluation, or to construct the solution set with linear lines with post-processing operations.

However, instead of dealing with post processing to reach solution sets consisting of linear lines, it is possible to reach the solution with different element and mesh types. On the other hand, in order to make an accurate analysis, constructing a model from surface elements, as used in the simulation in Figure 1, may require many additional measures such as excluding some unnecessary stiffnesses and other unnecessary properties from the simulation that are not needed at the core of the problem. In addition, it is certain that the process of increasing the mesh density to increase the precision of the calculation will increase the computational cost. But mesh density incrementation with element types low computational cost may offer a more efficient approach. However, it may not be possible to realize this approach by remaining in the option pool of the surface elements.

An approach in which the shortest path on a surface can be obtained as linear lines using element types with low computational cost is presented below. This approach is implemented by using truss elements that can rotate freely around the connection points instead of shell or membrane elements (Figure 2(b)) in modeling the surface of interest (Figure 2(c)).

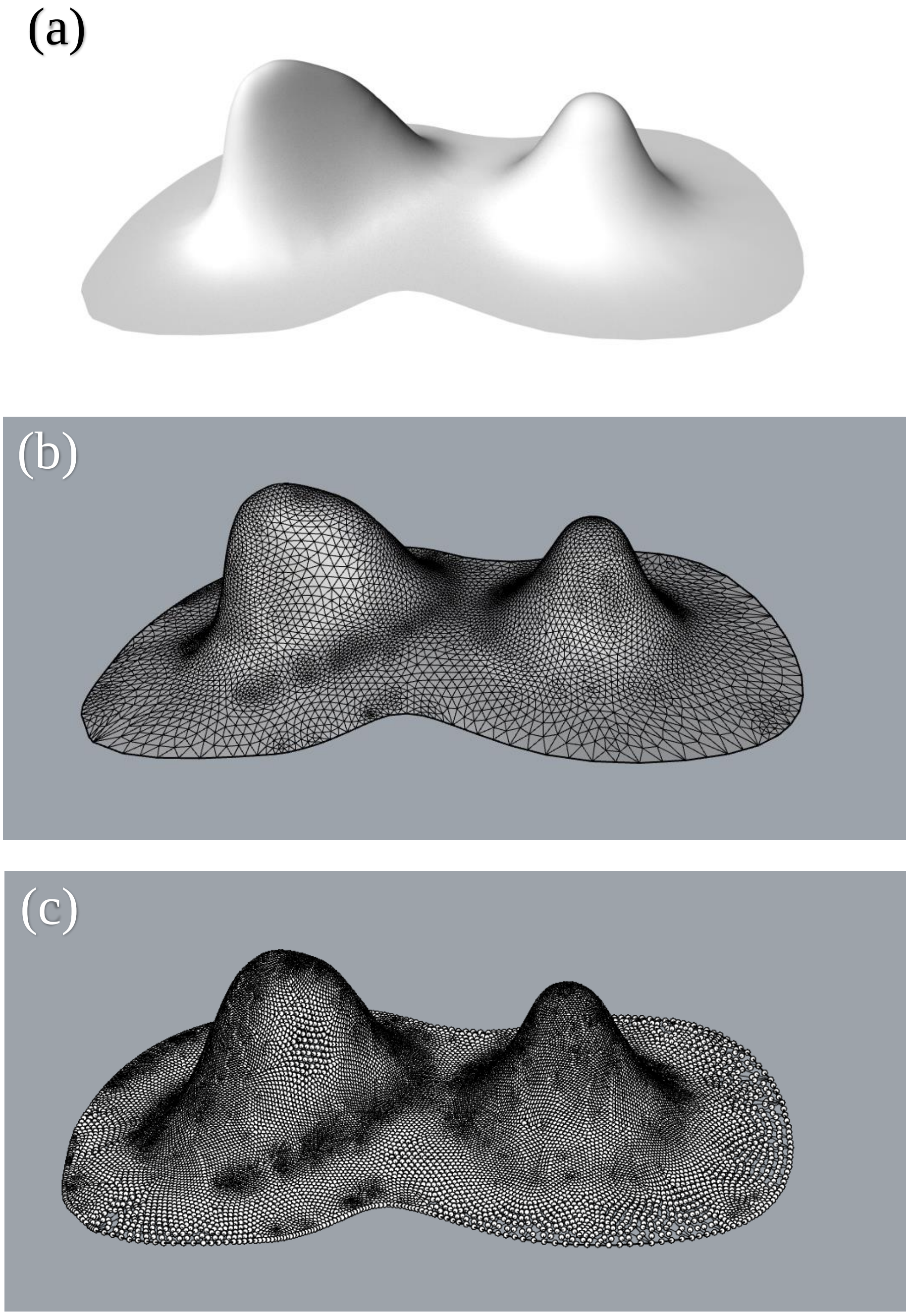

**Figure 2: (a)** Real surface, **(b)** Discrete surface mesh with triangular elements, **(c)** Transformed surface consisting of truss elements

With this approach, problems not only enable different element types of mechanical analogies to be solved, but also turn into a problem type that can be expressed and solved with discrete mathematics and graph theory. However, to speak again within the scope of mechanical analogies, there is an additional improvement that needs to be made in order for the system to give appropriate results in the case of converting the surface (Figure 3(a)) into a truss system (Figure 3(b)).

The exclusion of rigid triangular structures, which are likely to be found in truss systems obtained from surface triangles, by adding additional hinges is important in terms of making a suitable mechanical simulation [146]. In this context, joints should be added to the systems in order to eliminate rigid triangular structures. For this purpose, as can be seen in Figure 3(c), each truss is divided into two with a joint at the middle of the element [145].

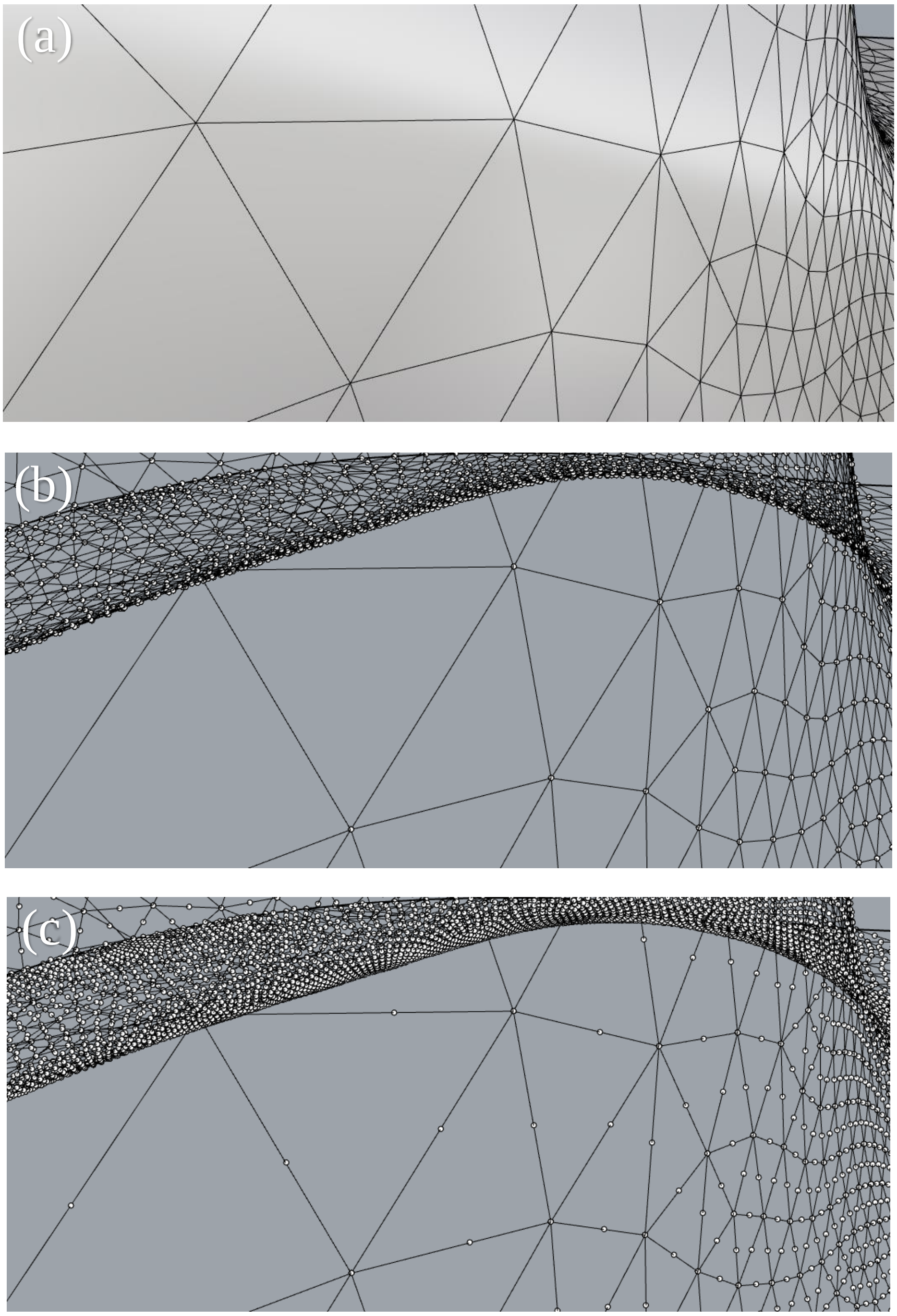

**Figure 3: (a)** Surface with triangular elements, **(b)** Surface demonstrated by truss elements, **(c)** Surface demonstrated by divided truss elements

No matter what type of different element types (triangular, quadrilateral, honeycomb etc.) the surfaces are represented, the surface meshes can be transformed into truss meshes by preserving the edges of surface elements . Each of the aforementioned approaches has advantages in different aspects, these advantages and disadvantages may also vary depending on problem types.

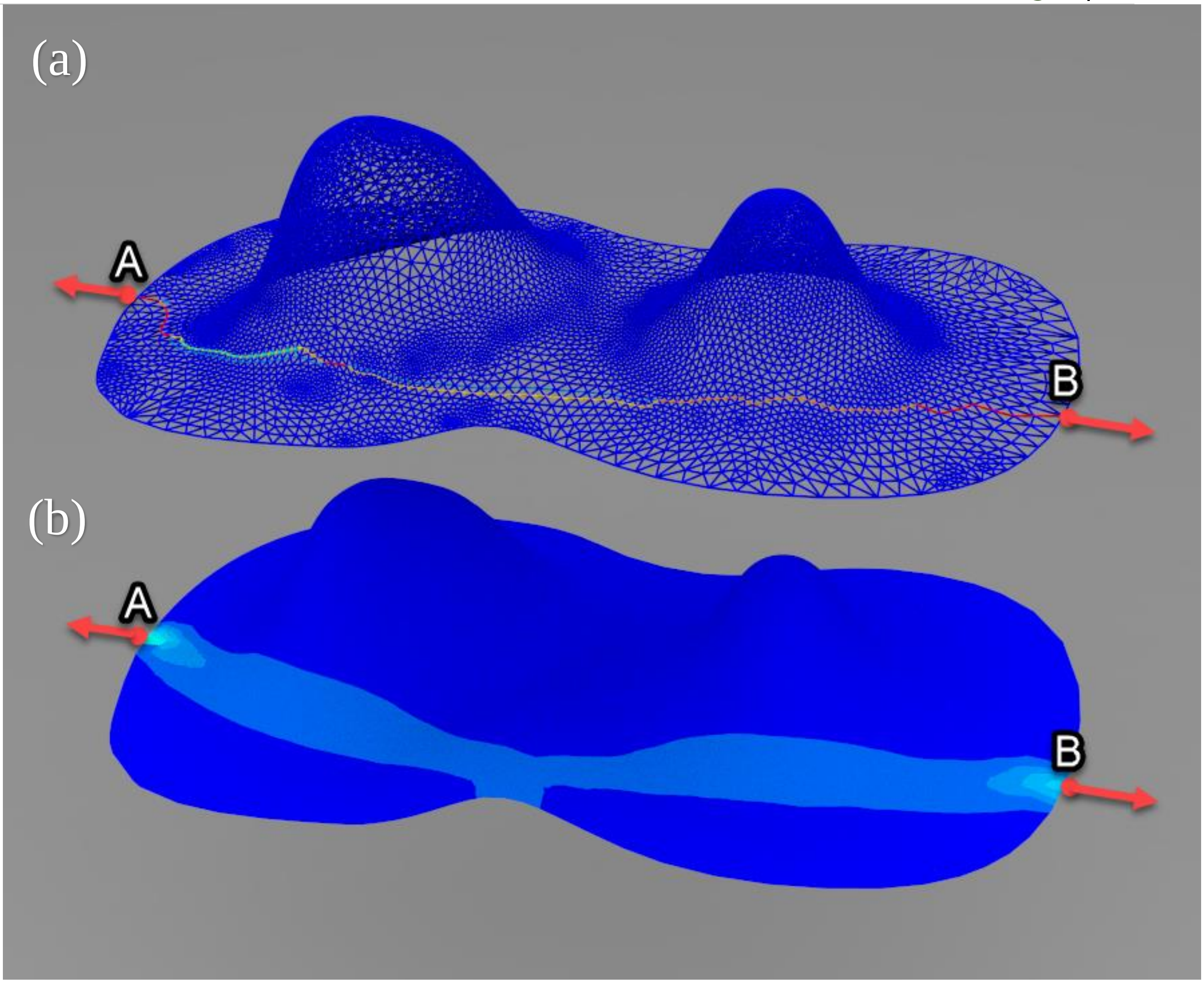

**Figure 4:** Stress results of analogical models (undeformed presentations), **(a)** Model with discretized truss elements, **(b)** Model with membrane elements (model has same number of nodes of model (a))

Looking at the solutions in Figure 4, the differences between truss elements (each divided) (Figure 4(a)) and solutions modeled with membrane elements (Figure 4(b)) can be easily seen. While the lines are obtained as the solution set in the system composed of truss elements, a certain region emerges as the solution set in the analysis made using the surface elements. Sharp turns seen in solution sets consisting of linear lines and large area coverages seen in solution sets obtained by using surface elements are features that can be arranged with post-processing if necessary in both approaches. Even increasing the number of elements used in truss systems and changing the color scale of the solution set revealed by the surface elements can also provide a distinctive observation of the solutions on the land.

***Mesh Conversion Procedure***

Any surface or volume meshes with point and edge definitions can be converted to a meshes with truss elements. However, it is not always possible to make this conversion, which also brings about a change in the mesh topology, directly with standard drawing programs or mesh editors. An algorithm (Figure 5) created in Grasshopper 3D visual programming language [147] was used to obtain the transformed meshes mentioned in this study. With this algorithm, the surface meshes were deconstructed based on their points and edges, and they were reconstructed in the desired format and brought together.

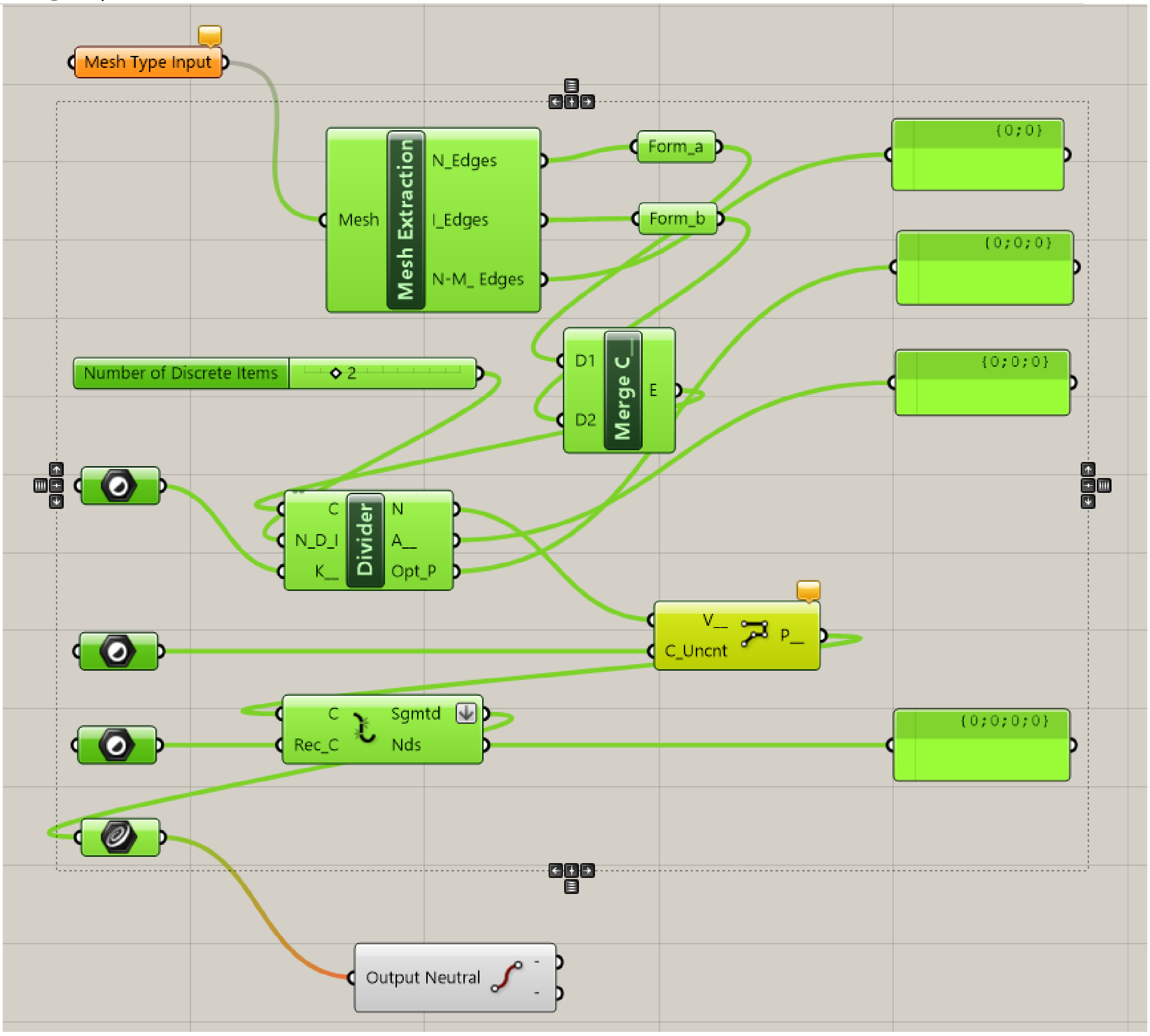

**Figure 5:** Visual algorithm for mesh conversion in Grasshopper 3D

The Visual Algorithm created in Grasshopper 3D, which is used for converting surface meshes (in file formats as STL (Standard Tessellation Language), OBJ (Wavefront), VRML (Wavefront) formats) to meshes consisting of divided truss elements and neutral file formats (such as IGES (Initial Graphics Exchange Specification)) is presented in Figure 5. Of course, the file types that can be processed and the approaches that can be used in conversions are not limited to those used in the study. Surface meshes (Figure 6 (b)) formed in various forms (Figure 6(a)) can be easily converted into meshes made of divided truss elements (Figure 6(c)) with the algorithm developed for this study. Although the possibility of forming rigid structures in the process of converting forms other than triangular surfaces into truss elements is not very likely, truss elements are presented by being discretized again in the images in Figure 6. In other words, regardless of whether the surface elements are quadrilaterals or triangulars, they are transformed into meshes (Figure 6(c)) consisting of discretized truss elements with the Grasshopper 3D algorithm presented in Figure 5.

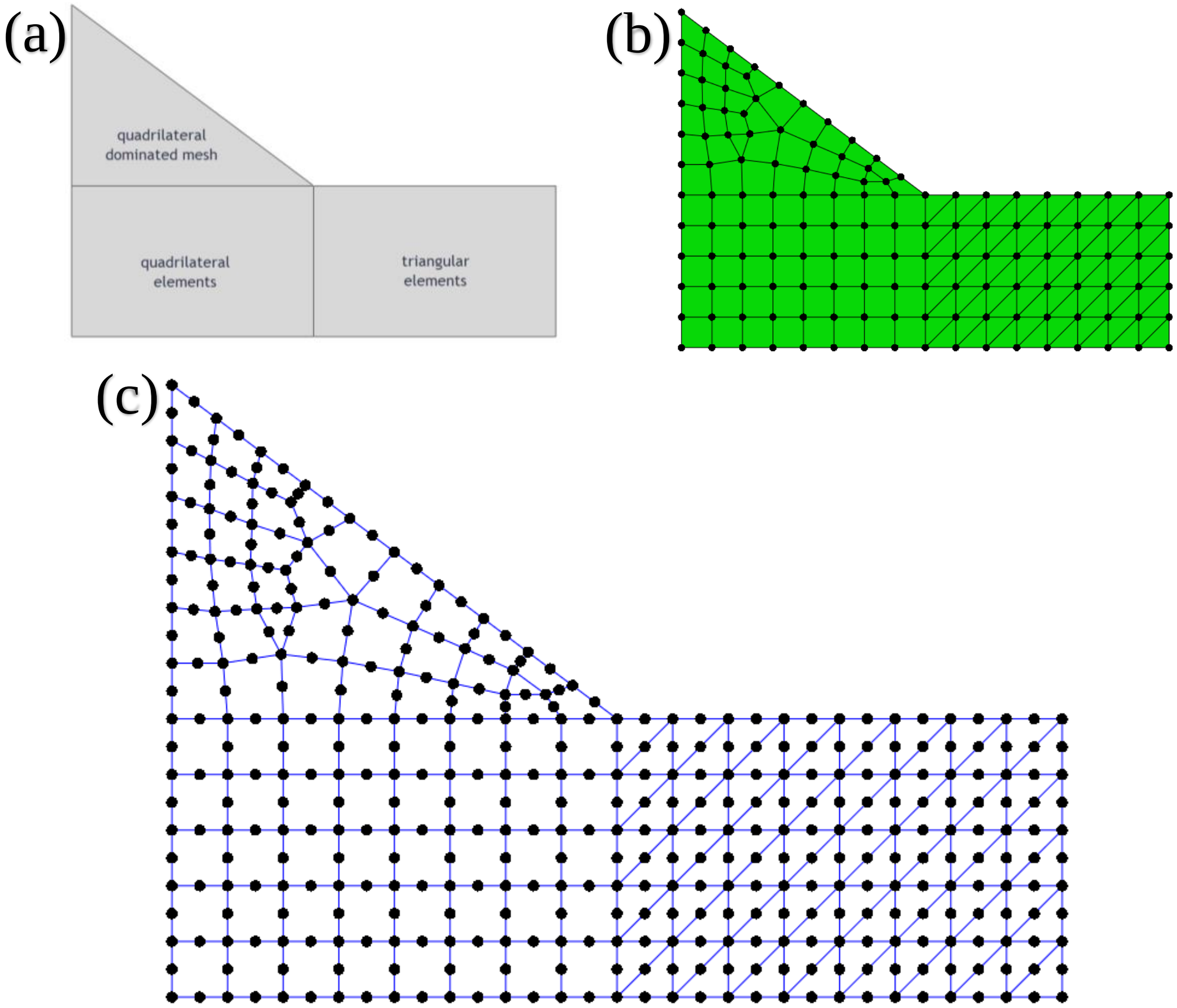

**Figure 6:** Element conversion, **(a)** Surface mesh generation fields for various strategies, **(b)** Generated surface meshes, **(c)** Mesh of discrete truss elements converted from surface meshes

***Comparison of meshes consisting of truss elements***

In the case of surface elements, it can be said that the increase in mesh intensity in shortest path analyzes is important in terms of obtaining more sensitive results. However, when it comes to truss element meshes derived from surface element meshes, increasing the number of truss elements by dividing has no effect on the solution of the shortest path problem, except for the elimination of rigid triangular structures, due to the preservation of the nodal points from the surface meshes. Unlike the evaluation of a pure mechanical problem, analyzes using models of truss element meshes require increasing the intensity of the surface meshes from which the existing truss elements are created, rather than dividing them within themselves, to produce more accurate results.

Truss elements can offer solutions with lower computational costs than other element types, but as seen in Figure 4(a), the path they offer as solutions are directly dependent on the mesh, and the path in the resulting solution set may contain sharp direction changes. Increasing mesh intensity or using higher number of truss elements to define the original surface with various methods can reduce the direction changes in the solution sets, or by using post-processing techniques on original geometry paths can contain more realistic lines.
In this context, it will be useful to compare the solutions offered by meshes consisting of truss elements created with several different approaches. In the sample prepared for this aim, there is a small hill on the surface where the shortest path between the points at its two opposite corners is desired to be determined, and the solutions obtained for the different meshes of the surface are presented in Figure 7. Structured meshes with full and partial orthogonal properties were used in the first two solutions, and unstructured meshes with different intensities were used in the others. In Figure 7(a), the lengths of the elements in the mesh consisting of truss elements derived from surface with quadrilateral elements formed entirely of orthogonal grids in the plan differ only in the hill region. Theoretically, the lengths of the paths consisting of orthogonal lines between two points in the plane are the same, but the lengths of the paths are slightly longer due to the slope only in the region around the hill and its immediate vicinity. For this reason, it is not possible to obtain a single shortest path in the analysis made by stretching the numerical model between two points.

In Figure 7(b), a surface mesh with a similar orthogonal grid formation in plan and consisting of isosceles right triangle elements was transformed into a mesh of truss elements. Since the orientation of the diagonal truss elements, which were not available in the previous system, is parallel to the direction connecting the two points, it was quite easy to determine the shortest path. However, if the additional diagonal elements were in the direction opposite to the current orientation, it would not have been possible to obtain the results so well.

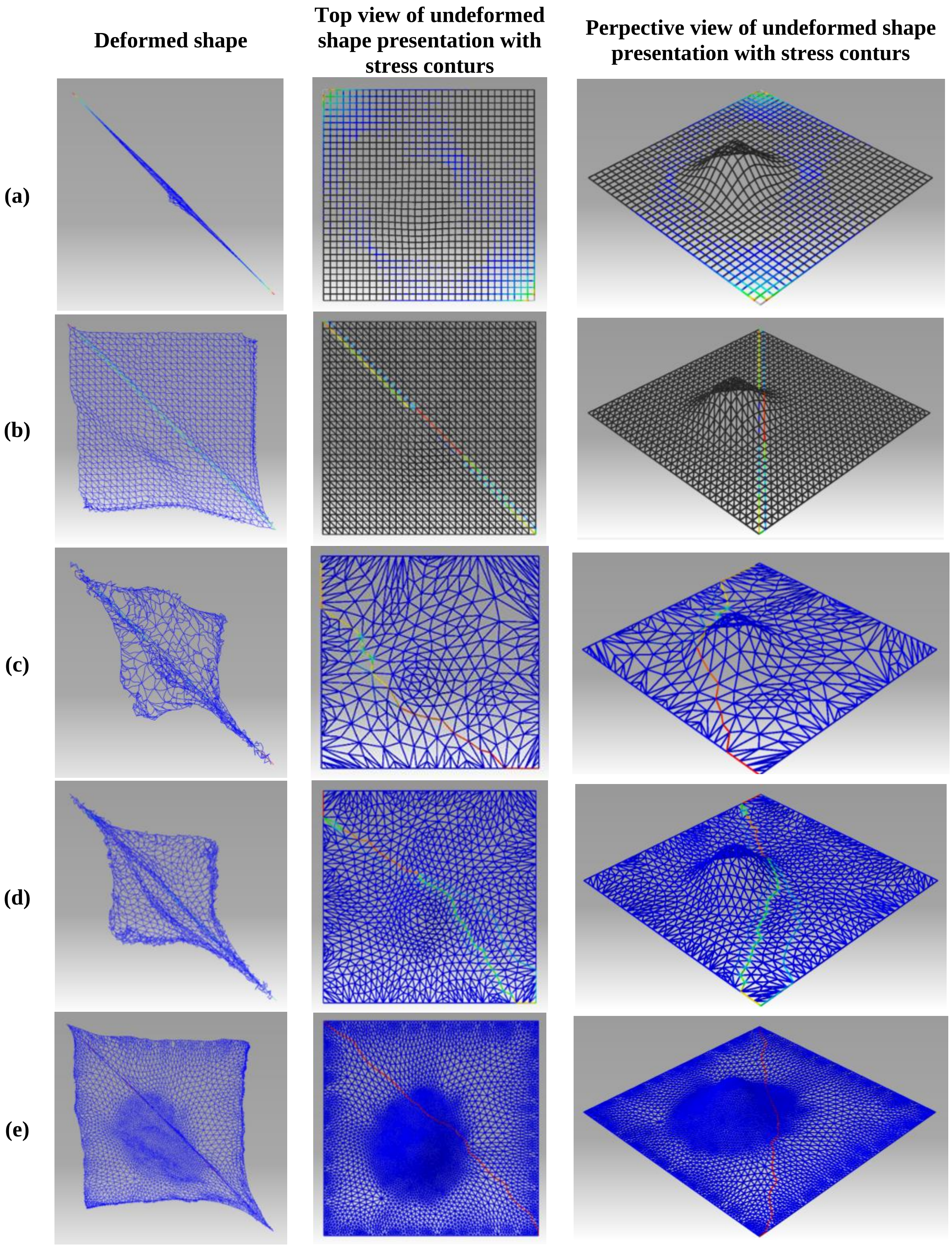

**Figure 7:** Shortest path solutions obtained from models with different mesh types

In Figure 7(c), (d) and (e), the naturalness of the solution paths presented by the models with different densities of unstructured meshes and their suitability to the original terrain increased depending on the increase in mesh intensity. Regardless of mesh intensity in the limitations of above meshes, it can be said that the performance of unstructured meshes is generally better in this problem. It should not be forgotten that there are mesh formations produced with very different techniques, apart from the meshes presented in Figure 7.

## Geographical application examples

The applications that were presented as an example problem and that wanted to determine the shortest distance between two points in a certain geographical region were also presented within the scope of the study. In the first example below, in the determination of the shortest path, a solution was obtained as if dealing with a geometry problem without considering obstacles, slopes, etc. that cannot be crossed for any vehicle.

In the first example, the points where the shortest path between them is desired are located on both sides of a valley, as seen in Figure 8 (a). The finite element model of the terrain, based on geographical data, consists of triangular surface elements (Figure 8(b)). The solution region was determined based on the stress, deformation and position values that occur by moving the points of interest of the finite element mesh away from each other (Figure 8 (c)). Any of these evaluation criteria can be used alone as evaluation criteria in flexible systems. The obtained shortest path solution set is presented in Figure 8(d) on the undeformed finite element mesh. Since surface elements are used in the analysis, a region was found as the solution set, and if desired, the region can be narrowed by refinement of the stress scale until it turns into a thin line or by a more direct post-processing approach.

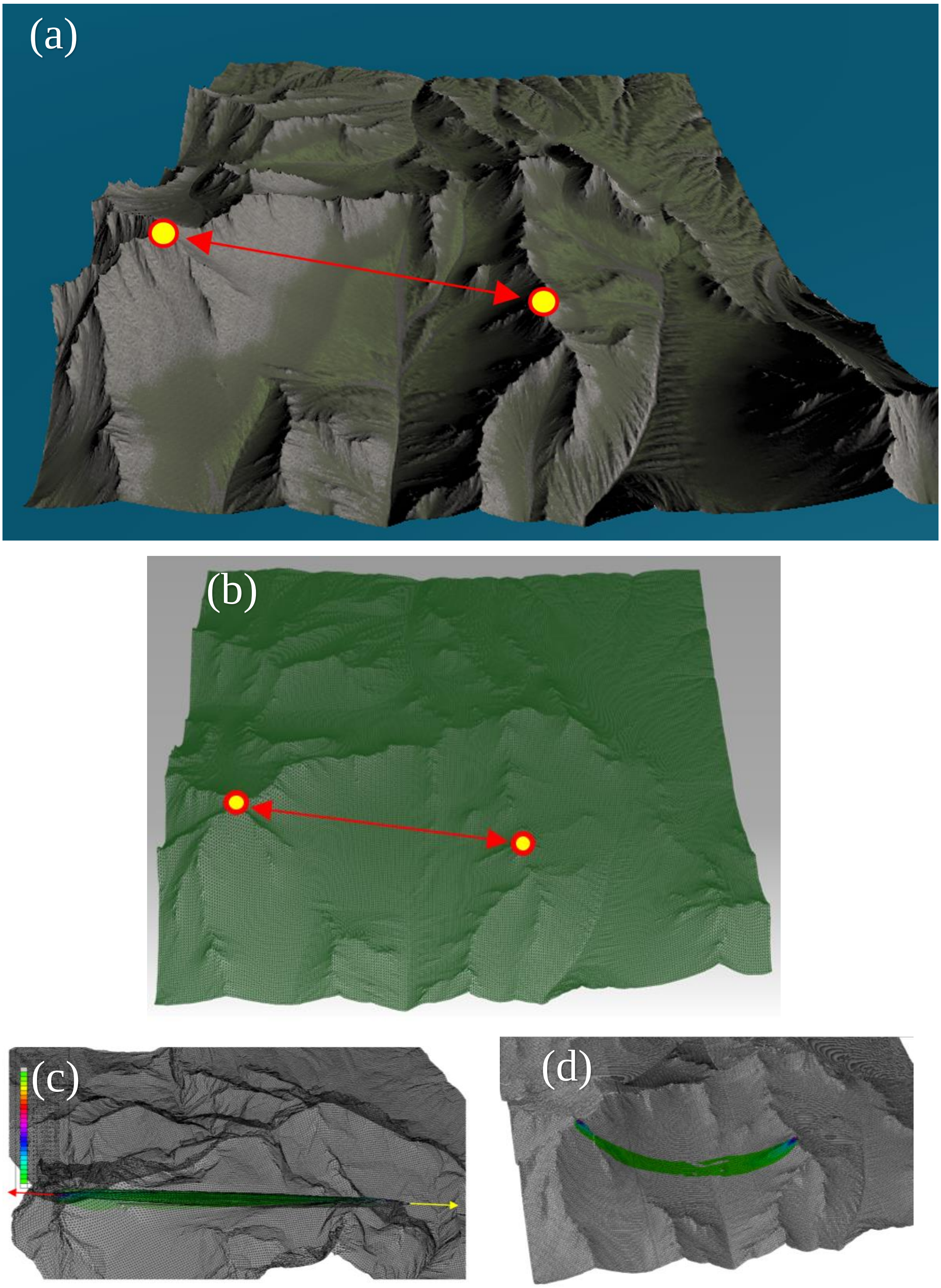

**Figure 8:** shortest path solution on terrain by using surface elements

In Figure 9, although the terrain and points of interest in Figure 8 are the same, in the analysis to be made in this part, some restrictions were created for the route that the shortest path between two points will pass through. Unlike the situation in Figure 8, where the shortest path region can be found predictably easily over a direct geographic connection, as in Figure 8, this example is set to show the constraints of the regions where the route can pass during the determination of the shortest path and to present the solution that the proposed approach can reveal. In this context, the terrain was flooded up to a certain height and the area created by the regions that could be used in the solution was reduced. Similar to the explanation above, an additional example of the practical applications of reducing the area that the route can pass through is the situation where an unmanned land vehicle will encounter, such as certain slopes, swampy and bush areas, which impede the progress of the vehicle.

With the proposed approach, the method to be used to solve such a problem is to remove the regions that cannot be used as a route from the finite element mesh and use the remaining part in the analysis. The two points in question are marked in Figure 9(a) and the solution is based on the finite element model created by using membrane elements. For this purpose, as in the previous examples, the terrain was modeled with membrane elements, allowing it to behave like a fabric, and the points were separated from each other until a tense and linear line was formed between the two points, as seen in Figure 9(b). In Figure 9(b) the final stress state of the simulation is superimposed on the original geometry of the terrain (Figure 9(c)) to see the placement of the line on the ordinary state of the terrain, which may not be clearly distinguishable on the deformed shape (Figure 9(c)). Due to the element type used, a region with a certain area connecting the two points is determined as the region through which the shortrest path will pass, and if desired, the said region can be further narrowed as mentioned before.

In Figure 9(d), it is determined by using a finite element model created with truss elements, as in Figure 4(a) and Figure 7, in order to determine the desired route with lower computational cost and linear lines instead a region. As it can be seen more clearly from Figure 9(e), which is the close view image, since the finite element mesh is created with the structured mesh technique, the solution line found, as in the previous examples, is highly affected by the finite element mesh and is somewhat incompatible with the natural surface of the land. This situation is clearly seen in Figure 7 in the study in which different meshes are compared. In this context, additional analyzes were made using a model with a finite element mesh created with the unstructured mesh technique in the region of interest. As can be clearly seen from Figure 9(d) and Figure 9(e), the regions that cannot be used as route region were removed from the images in order to focus on the results of the analyzes made using unstructured mesh (Figure 9(f), Figure 9(g)). When looking at the results for the region, the close view of which is presented in Figure 9(f) and the close view in Figure 9(g), although fewer elements are used, more suitable routes to the land surface were obtained compared to the analyzes made using structured mesh.

As a striking point in the results, in Figure 9(d) and Figure 9(e), three alternative routes stand out in the region on the left side of the land, and the middle one of these three routes offers the shortest path among them, due to the fact that it has the biggest stress and deformation (not shown numerically). However, since there is a very small distance difference between the other two alternative routes, they have not been removed from the images in order to be seen as options that can be used on alternative routes for different reasons.

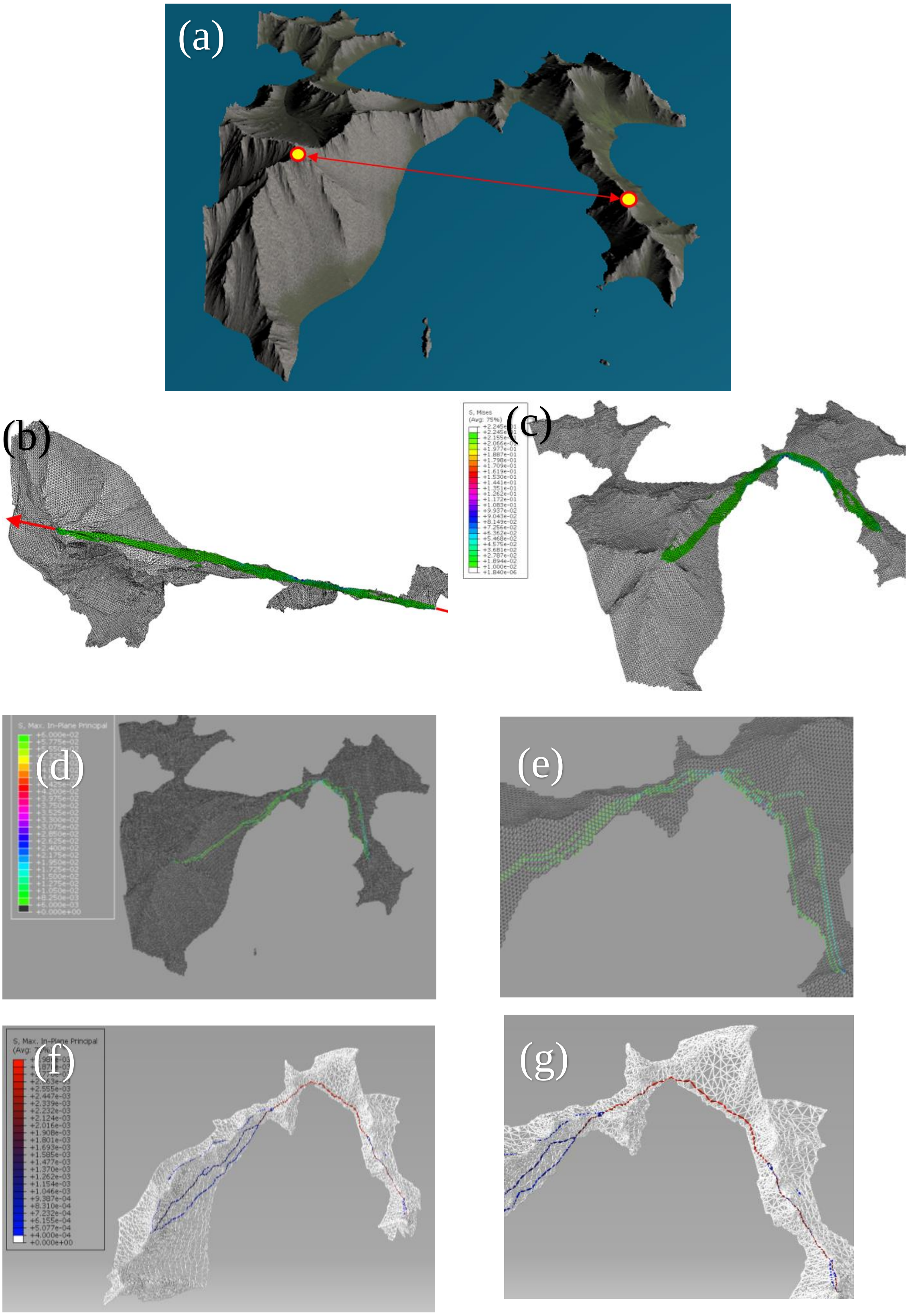

**Figure 9:** Shortest path solution on terrain by various elements and mesh structures

## Results

In this study, various solutions based on mechanical analogies were proposed to find the shortest path between two points on a surface. The finite element method was used in realizing the mechanical analogies of flexible systems, and the results obtained by using various element types and meshes in the creation of the finite element models used in the solutions were compared.

In analyzes using surface elements, there is a solution region through which the shortest path must pass, depending on the surface properties, the mesh intensity, and the threshold values of the discriminating parameters used to determine the solution set. Regional solution sets can be easily reduced to a linear lines with post processing operations, but this study focuses directly on the results of primary analysis.

Another element type used in the analysis is truss elements, and it can be thought that the computational cost of truss elements is lower than the surface elements. Meshes of truss elements can be converted from various surface mesh types or constructed from data from various sources.

The results obtained show that the results obtained from the models in which truss elements are used in the analyzes related to the determination of the shortest paths are highly affected by forms of the meshes. In the comparative analyzes made in this context, it can be said that unstructured meshes generally perform better than structured meshes, based on the samples within the scope of the study.

Although primary analyzes are included in this study, it is clear that solution sets consisting of linear lines obtained by using truss elements can be made more suitable for the land by post-processing. It is obvious that the sensitivity will increase with the increase of mesh intensity, valid for both the meshes consisting of truss elements and surface elements. It should be noted that the increase in the intensity of mesh consisting of truss elements depends on the increase in the mesh intensity of the surfaces from which they are derived. Both models, in which surface elements and truss elements are used, have their own advantages and it is possible to choose the most suitable one according to the application area. As suggested in Altintas [144], the interaction property should not be used in the mechanical analogical model unless it is part of the original problem. In this way, constraints that are not contained in the original problem are not included in the problem. In the analyzes made with models created using truss elements, they were subjected to additional discretization process in order to avoid the emergence of rigid triangular structures that could hinder the appropriate results from being obtained [146].

The presented models, and especially the models consisting of truss elements, can be solved very quickly not only with the finite element method, but also with simulations based on rigid body dynamics (connected multiple rigid bodies etc.). In fact, such a procedure can theoretically be much faster and more effective than the analysis used with the finite element method, but in this study, the finite element method was used in order to include more mechanical parameters in the use of mechanical analogies and to understand the subject on a mechanical basis. On the other hand, it is clear that the meshes consisting of truss elements formed by the finite element method can also be solved by the methods of Graph theory. Considering the issue as an optimization problem, the problem can be solved with a similar approach based on mechanical analogy [146].

Finally, since the approaches presented in this report can be used with almost all current finite element based engineering software, it is thought to be valuable not only theoretically but also for practitioners.

MCBU CIVIL ENGINEERING REPORTS

# Finding Shortest Path on a Terrain Surface by Using Finite Element Method

Gökhan Altıntaş

MCBUCIVILENG.R-2022.1

SCIENCE + TECHNOLOGY